\documentclass[letterpaper,english,aps, showpacs,showkeys]{revtex4-1}
\usepackage[T1]{fontenc}
\usepackage{color}
\usepackage{textcomp}
\usepackage{amsmath}
\usepackage{amssymb}
\usepackage{graphicx}

\makeatletter

\providecommand{\tabularnewline}{\\}

\@ifundefined{showcaptionsetup}{}{%
 \PassOptionsToPackage{caption=false}{subfig}}
\usepackage{subfig}
\makeatother

\usepackage{babel}
\begin{document}

\title{The statistical properties of q-deformed Morse potential for some diatomic molecules via Euler-MacLaurin
method in one dimension}

\author{Abdelmalek Boumali}

\affiliation{Laboratoire de Physique Appliqu\'ee et Th\'eorique, ~\\
Universit\'e Larbi T\'ebessi -T\'ebessa-, 12000, W. T\'ebessa, Algeria.}
\email{boumali.abdelmalek@gmail.com}

\begin{abstract}
In this paper, we present a closed-form expression of the vibrational partition function for the one-dimensional q-deformed Morse potential energy model. Through this function, the related thermodynamic functions are derived and studied in terms of the parameters of the model. Especially, we plotted the q-deformed vibrational partition function and vibrational specific heat for some diatomic molecule systems such as $\text{H}_{2}$, HCl, LiH, and CO. The idea of a critical temperature $T_{C}$ is introduced
in relation to the specific heat.
\begin{description}
\item [{PACS~numbers}] 03.65.Ca; 03.65.Fd.{\small \par}
\end{description}
\end{abstract}

\maketitle

\section{Introduction}

Quantum groups and quantum algebras have attracted much attention of physicists and mathematicians during the last eight years. There had been a great deal of interest in this field, especially after the introduction of the q-deformed harmonic oscillator. Quantum groups and quantum algebras have found unexpected applications in theoretical physics \cite{1}. From the mathematical point of view, they are q-deformations of the universal enveloping algebras of the corresponding Lie algebras, being also concrete examples of Hopf algebras. When the deformation parameter q is set equal to 1, the usual Lie algebras are obtained. The realization of the quantum algebra SU(2) in terms of the q-analogue
of the quantum harmonic oscillator \cite{2,3} has initiated much work on this topic \cite{4,5,6}. Biedenharn and Macfarlane \cite{2,3} have studied the q-deformed harmonic oscillator based on an algebra of q-deformed creation and annihilation operators. They have found the spectrum and eigenvalues of such a harmonic oscillator under the
assumption that there is a state with the lowest energy eigenvalue. The method of q-deformed quantum mechanics was based on the Heisenberg commutation relations for bosons. The main parameter of this method is a real number $q\in [0,1]$, called deformation.\par
Recently, the theory of the q-deformed has become a topic of great interest in the last few years and it has been finding applications in several branches of physics because of its possible applications in a wide range of areas, such as a q-deformation of the harmonic oscillator \cite{7}, a q-deformed Morse oscillator \cite{8}, a classical and quantum q-deformed physical systems \cite{9}, Jaynes-Cummings
model and the deformed-oscillator algebra \cite{10}, q-deformed super-symmetric quantum mechanics \cite{11} for some modified q-deformed potentials \cite{12}, on the Thermo-statistic properties of a q-deformed ideal Fermi gas \cite{13}, q-Deformed Tamm-Dancoff oscillators\cite{14}, q-deformed fermionic oscillator algebra and thermodynamics \cite{15}, Algebraic Approach to Thermodynamic Properties
of Diatomic Molecules\cite{16}, the quantum group symmetry of diatomic molecules \cite{17}, and finally on the fermionic q deformation and its connection to the thermal effective mass of a quasi-particle \cite{18}.\par
In the atomic and molecular physics, the interaction between atoms in diatomic and even in polyatomic molecules is usually described  by the Morse potential. This potential is the most simple and realistic anharmonic potential model, which has been widely used in the description of vibrational motion of diatomic molecules. In algebraic approach, the Morse potential can be written in terms of the generators of SU(2) group. The quantum relation between q-deformed harmonic oscillator and the Morse potential was considered, and the extended SU (2) model (q-Morse potential) has been also developed to compare with phenomenological Dunham expansion and experimental data for numbers of diatomic molecules \cite{16,17}.  Analytical representations of thermodynamic functions of gases over the whole temperature range from zero to the thermal dissociation limit have aroused much interest in dealing with diatomic and polyatomic systems. Through the exact form of its spectrum of energy, the vibration partition function, which is of great importance to many issues in chemical physics and engineering, can be obtained. Following its definition, the molecular vibrational partition function can be calculated by direct summation over all possible vibrational energy levels. Many efforts have been made to acquire explicit expressions of partition functions for molecular potential energy models in diatomic molecules and polyatomic molecules: in this way, and in the non-relativistic case, the investigation of thermodynamic functions of some type of potentials such as Morse and improved Manning-Rosen potentials, improved Rosen-Morse and Tiez oscillators, through a partition function and its derivatives with respect to temperature were an important field of research in the literature \cite{19,20,21,22}: Strekalov \cite{19}, by using the Poisson summation formula, found an accurate closed-form expression for the partition function of Morse oscillators. Following the same method, the authors \cite{20,21,22} have been calculated the thermal properties of some diatomic molecules for the different type of potentials.\par
The aim of this paper is to calculate the vibrational temperature thermodynamics of diatomic molecules and to obtain the basic thermodynamic functions in terms of the parameter of deformation $q$ via Euler-MacLaurin method \cite{23}. The reason is twofold: (i) in our best knowledge, the study of these thermal properties by varying the parameter of deformation $q$ does not exist in the literature, and (ii) consequently, in the same context, we will propose another method based on the Euler-Maclaurin formalism to determine these properties: this method of calculation, as far as we know, is used for the first time for this genre of study. Thus, this paper can be considered as a first step of incorporating the deformation and the finite number of bound vibrational states into the thermodynamic description of molecular systems.\\
The rest of the paper is organized as follows. In Section II, we review the eigensolutions of one-dimensional q-deformed Morse potential. In Section III, we derive a q-deformed Morse vibrational partition function, and consequently obtain, through this function the basic thermodynamic functions, such as the mean vibrational energy,
specific heat, and free energy. Section IV will be a conclusion.

\section{Eigensolutions of one-dimensional q-deformed Morse potential: review}

\subsection{The q-deformed Morse potential in one-dimension Schrödinger equation}

The time independent Schrödinger equation with the q-deformed Morse
potential $V(x)$ where 
\begin{equation}
V\left(x\right)=V_{0}\left(e^{-2\alpha x}-2qe^{-\alpha x}\right),\label{eq:1}
\end{equation}
can be written as follows 
\begin{equation}
\left[\frac{d^{2}}{dx^{2}}-\frac{2m}{\hbar^{2}}\left\{ V_{0}\left(e^{-2\alpha x}-2qe^{-\alpha x}\right)-E_{n}\right\} \right]\psi\left(x\right)=0.\label{eq:2}
\end{equation}
Here $V_{0}$ is the dissociation energy , $\alpha$ is the range
parameter and the parameter $m$ is the reduced mass of the molecule.
The eigensolutions of Eq. (\ref{eq:2}) are well discussed in the
literature, and following \cite{24}, they are written as follows:
\begin{equation}
E_{n}=-\frac{\hbar^{2}\alpha^{2}}{2m}\left(\mu-n-\frac{1}{2}\right)^{2},\label{eq:3}
\end{equation}
\begin{equation}
\psi_{n}\left(x\right)=\sqrt{\frac{n!}{\left(2n+s\right)!}}\rho^{s}e^{-\frac{\rho}{2}}L_{n}^{2s}\left(\rho\right),\label{eq:4}
\end{equation}
with
\begin{equation}
L_{n}^{k}\left(\rho\right)=\frac{\Gamma\left(k+n+n\right)}{n!\Gamma\left(k+1\right)}\,_{1}F_{1}\left(-n,k+1,\rho\right),\label{eq:5}
\end{equation}
is generalized Laguerre polynomials and $_{1}F_{1}$ is hypergeometric
functions.

Now, by using the following substitutions
\begin{equation}
s=\mu-n-\frac{1}{2},\label{eq:6}
\end{equation}
\begin{equation}
\mu=\frac{\nu q}{2}\,\text{and}\,\nu^{2}=\frac{8mV_{0}}{\hbar^{2}\alpha^{2}},\label{eq:7}
\end{equation}
and Eq. (\ref{eq:7}), Eq. (\ref{eq:3}) becomes
\begin{equation}
E_{n}=-q^{2}V_{0}\left(1-\frac{n+\frac{1}{2}}{\mu}\right)^{2}.\label{eq:8}
\end{equation}
Expression (\ref{eq:8}) is the q-deformed vibrational energy levels.
The maximum value $n_{max}$ can be obtained by setting $\frac{dE_{n}}{dn}=0,$
so we have
\begin{equation}
n_{max}=\frac{\nu q}{2}-\frac{1}{2}.\label{eq:9}
\end{equation}
The $n_{max}$ means the upper bound vibration quantum number.

\begin{figure}
\includegraphics{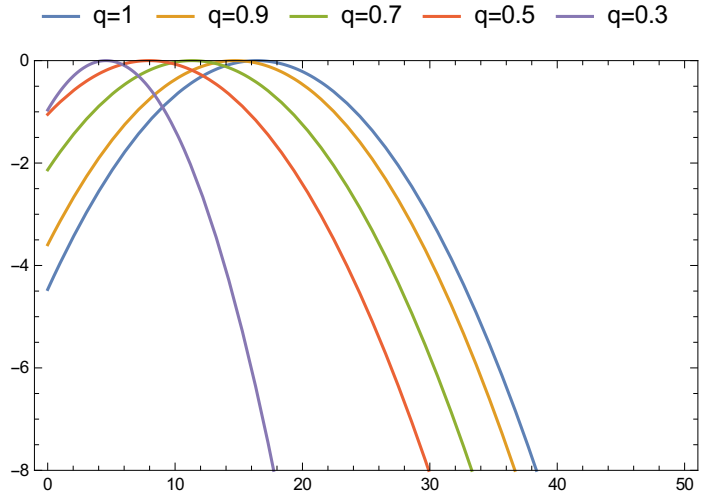}

\caption{ \label{fig:1}The q-deformed vibrational energy level vs $n$ for
different values of $q$ with $0<q<1$}

\end{figure}

The Fig. \ref{fig:1}, shows the q-deformed vibrational energy levels vs $x$ for different values of $q$: following this figure, we observe that the maximum value $n_{max}$ decreases with the parameter of deformation until becoming $0$ for certain value of $q$. 
The case when $q=1$ leads to the usual form of the vibrational energy levels of the standard Morse potential \cite{25}.

\begin{figure}
\includegraphics{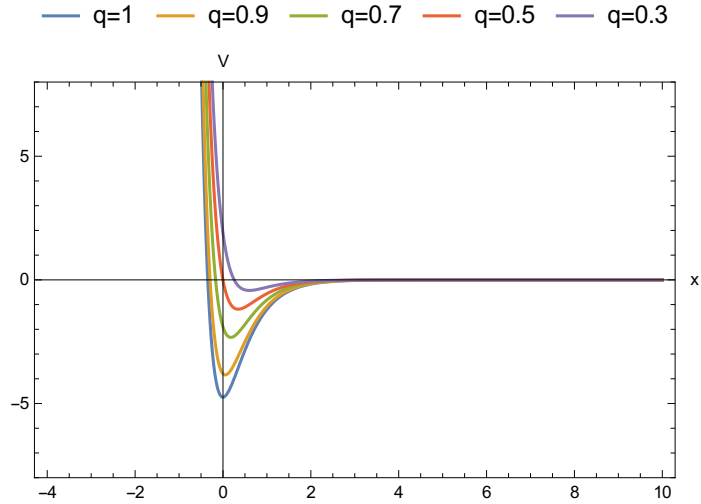}

\caption{\label{fig:2}Variation of q-deformed Morse potential $V\left(x\right)$
versus $x$ for different values of $q$ with $0<q<1$. }
\end{figure}

The Fig. (\ref{fig:2}) 
 presents the variation of q-deformed Morse potential vs $x$ for different values of $q$: the hosting of the number of quantum levels of the potential, decreases with q until becoming 0 for certain values
of q. The case when $q=1$ corresponds to the usual Morse potential.

In what follows, we focus on the influence of the parameter of deformation
on the thermal properties of the system in question obtained by using
the Euler-Maclaurin method. Especially, we choose to study only the
curves of vibrational partition function and the vibrational specific
heat for some diatomic molecules such as $\text{H}_{2}$, HCl, LiH
and CO most widely studied in the literature.

\section{Thermal properties of q-deformed Morse potential for some Diatomic
molecules in one dimension}

\subsection{Methods}

As we know, all thermodynamic quantities can be obtained from the partition function Z, therefore, the partition function of the system is the starting point to derive all thermal properties of the system
in question. The vibrational partition function can be calculated by direct summation over all possible vibrational energy levels available to the system. Given the energy spectrum, the partition function Z of the q-deformed
Morse potential at finite temperature $T$ is obtained through the Boltzmann factor is
\begin{equation}
Z=\sum_{n=0}^{\infty}e^{-\beta\left(E_{n}-E_{0}\right)}.\label{eq:10}
\end{equation}
\textcolor{black}{with $\beta=\frac{1}{k_{B}T}$ and with $k_{B}$
is the Boltzmann constant. Now in order to calculate the partition function, we use the Euler\textendash MacLaurin formula: according to this approach, the sum transforms to the integral as follows 
\begin{equation}
\sum_{n=0}^{\infty}f\left(x\right)=\frac{1}{2}f\left(0\right)+\int_{0}^{\infty}f\left(x\right)dx-\sum_{p=1}^{\infty}\frac{B_{2p}}{\left(2p\right)!}f^{\left(2p-1\right)}\left(0\right),\label{eq:11}
\end{equation}
where $B_{2p}$ are the Bernoulli numbers, $f^{\left(2p-1\right)}$
is the derivative of order (2 p \textminus{} 1). Up to $p=3$, and
with $B_{2}=\frac{1}{6}$ and $B_{4}=-\frac{1}{30}$ the partition
function $Z$ is written as
\begin{align}
Z & =e^{-\beta\left\{ q^{2}V_{0}\left(1-\frac{1}{q\nu}\right)^{2}\right\} }\sum_{n=0}^{n_{max}}e^{-\beta\left\{ -q^{2}V_{0}\left(1-\frac{2n+1}{q\nu}\right)^{2}\right\} }.\nonumber \\
 & =\frac{1}{2}+\int_{0}^{n_{max}}f\left(x\right)dx-\sum_{p=1}^{\infty}\frac{B_{2p}}{\left(2p\right)!}f^{\left(2p-1\right)}\left(0\right),\label{eq:12}
\end{align}
with 
\begin{equation}
f\left(x\right)=e^{-\beta\left\{ -q^{2}V_{0}\left(1-\frac{2n+1}{q\nu}\right)^{2}\right\} }.\label{eq:13}
\end{equation}
After some calculations, we arrive at the final form of the partition function $Z$ where
\begin{align}
Z & \left(\beta,q,a\right)=\frac{1}{2}+\frac{0.0233\beta q\left(1-\frac{0.0147}{aq}\right)}{a}\nonumber \\
 & -\frac{1}{720}\left(\frac{0.022\beta^{3}q^{3}\left(1-\frac{0.0147}{aq}\right)^{3}}{a^{3}}+\frac{0.007\beta^{2}q\left(1-\frac{0.0147834}{aq}\right)}{a^{3}}\right)\nonumber \\
 & -\frac{0.057\left(a-67.6436a^{2}q\right)e^{-\frac{4.7446\beta\left(0.0147834a-a^{2}q\right)^{2}}{a^{4}}}\text{erf}\left(2.17\sqrt{-\frac{\beta\left(0.0147a-a^{2}q\right)^{2}}{a^{4}}}\right)}{a\sqrt{\beta\left\{ q\left(\frac{0.00233}{a}-0.078q\right)-\frac{0.0000172}{a^{2}}\right\} }},\label{eq:14}
\end{align}
where erf denotes the usual error function and $a=\frac{1}{\alpha}$. }

In what follows, all thermodynamic properties of the one-dimensional q-deformed Morse potential, such as the free energy, the entropy, total energy and the specific heat, can be obtained through the numerical partition function $Z(q,\beta, a)$. These thermodynamic functions for the diatomic molecules system can be calculated from the following expressions
\begin{align}
F & =-\frac{1}{\beta}\ln Z,\label{eq:15}\\
U & =-\frac{d\ln Z,}{d\beta}\label{eq:16}\\
S & =\ln Z-\beta\frac{d\ln Z}{d\beta},\label{eq:17}\\
C & =\beta^{2}\frac{d^{2}\ln Z}{d\beta^{2}}.\label{eq:18}
\end{align}
In our study, we will concentrate only on the influence of parameter of deformation $q$ on vibrational partition function and vibrational specific heat.

\subsection{Applications for some diatomic molecules}

In the present work, we choose four diatomic molecules, $\text{H}_{2}$, HCl, LiH and CO which have been most widely studied in the literature. The typical values of molecular constants for the electronic state of these molecules are given in Table. \ref{tab:1} \cite{26,27}:
following this table we can find the reduced mass $m$, the depths of the potential well, (dissociation energies) $V_{0}$ and the widths of the potential well $a$. 
\begin{table}[h]
\begin{tabular}{|c|c|c|c|c|}
\hline 
Molecule & $m\left(\text{amu}\right)$ & $V_{0}\left(\text{eV}\right)$ & $\alpha\left(A^{-1}\right)$ & $a=\frac{1}{\alpha}\left(A\right)$\tabularnewline
\hline 
\hline 
HCl & \textbf{0.9801045} & \textbf{4.61907} & \textbf{2.38057} & \textbf{0.420067463}\tabularnewline
\hline 
$\text{H}_{2}$ & \textbf{0.50391 } & \textbf{4.7446} & \textbf{1.440558} & \textbf{0.694175451}\tabularnewline
\hline 
CO & \textbf{6.8606719 } & \textbf{11.2256} & \textbf{2.59441} & \textbf{0.385444089}\tabularnewline
\hline 
LiH & \textbf{0.8801221} & \textbf{2.515287} & \textbf{1.7998368} & \textbf{0.55560593}\tabularnewline
\hline 
\end{tabular}

\caption{\label{tab:1}Spectroscopic parameters of selected molecules, used in the present calculation.}
\end{table}

Before to determine the thermal properties of our problem in question, we first calculate the number of maximum quantum levels $n_{max}$:
for this we use that $\hbar c=1973.269\,\text{eV}\,\text{A}^{\text{\textdegree}}$
and $1\text{amu}=931.5\times10^{6}\text{eV}\,\left(\text{A}^{\text{\textdegree}}\right)^{-1}$.
These values are depicted in Table. \ref{tab:2}
\begin{table}[h]
\begin{tabular}{|c|c|c|c|c|}
\hline 
$q$ & $\text{H}_{2}$ & $\text{HCl}$ & $\text{LiH}$ & $\text{CO}$\tabularnewline
\hline 
\hline 
1 & \textbf{22} & \textbf{19} & \textbf{36} & \textbf{73}\tabularnewline
\hline 
0.9 & \textbf{20} & \textbf{17} & \textbf{15} & \textbf{66}\tabularnewline
\hline 
0.7 & \textbf{15} & \textbf{13} & \textbf{12} & \textbf{51}\tabularnewline
\hline 
0.5 & \textbf{11} & \textbf{9} & \textbf{8} & \textbf{36}\tabularnewline
\hline 
0.3 & \textbf{6} & \textbf{5} & \textbf{4} & \textbf{21}\tabularnewline
\hline 
\end{tabular}

\caption{\label{tab:2}The values of $n_{max}$ used in the calculations.}
\end{table}
: following this table, the number of quantum levels, in the well potential, are finite and decrease with the parameter of deformation $q$ until becoming $0$ for certain values of $q$.
With the aid of these values, the q-deformed vibrational partition function $Z$ are determinate, and consequently all thermal properties of our system can be found easily.

\begin{figure}
\subfloat[Both vibrational partition function $Z$ and specific heat $\frac{C}{k_{B}}$
of $\text{H}_{2}$ as a function of $\beta$ for different values
of $q$.]{\includegraphics[scale=0.6]{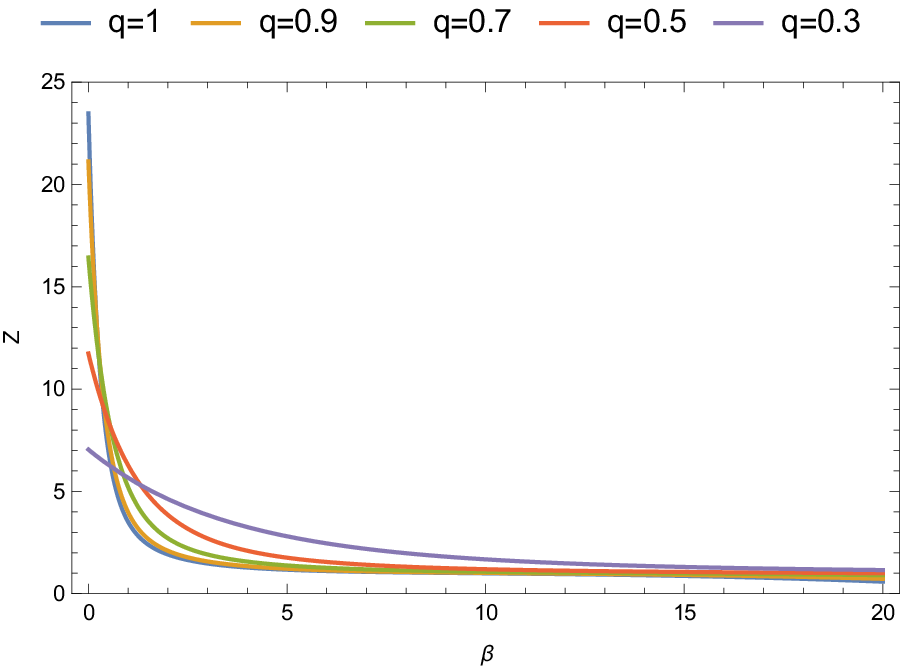}\includegraphics[scale=0.62]{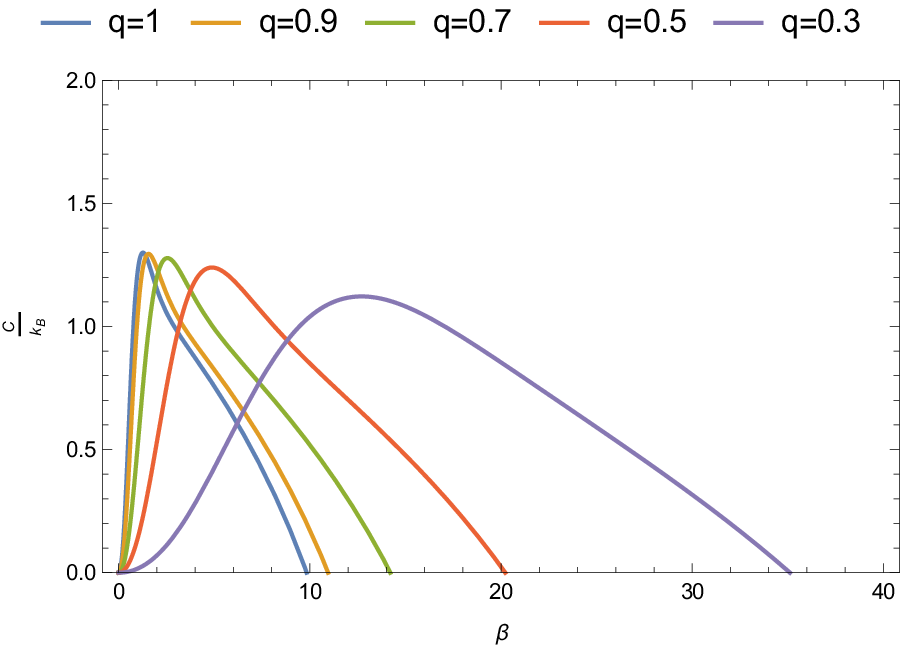}

}

\subfloat[Both vibrational partition function $Z$ and specific heat $\frac{C}{k_{B}}$
of $\text{HCl}$ as a function of $\beta$ for different values of
$q$.]{\includegraphics[scale=0.6]{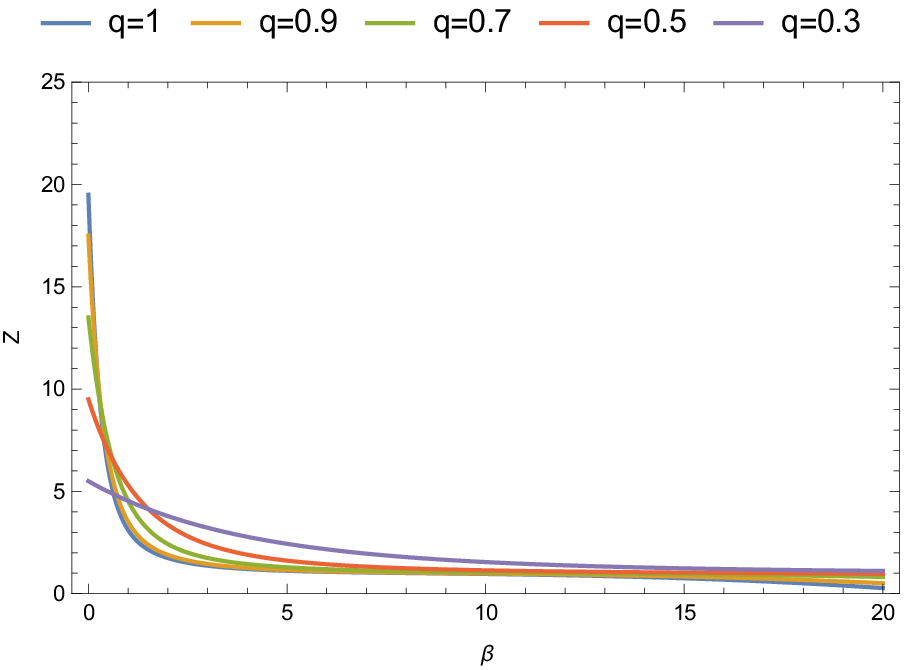}\includegraphics[scale=0.62]{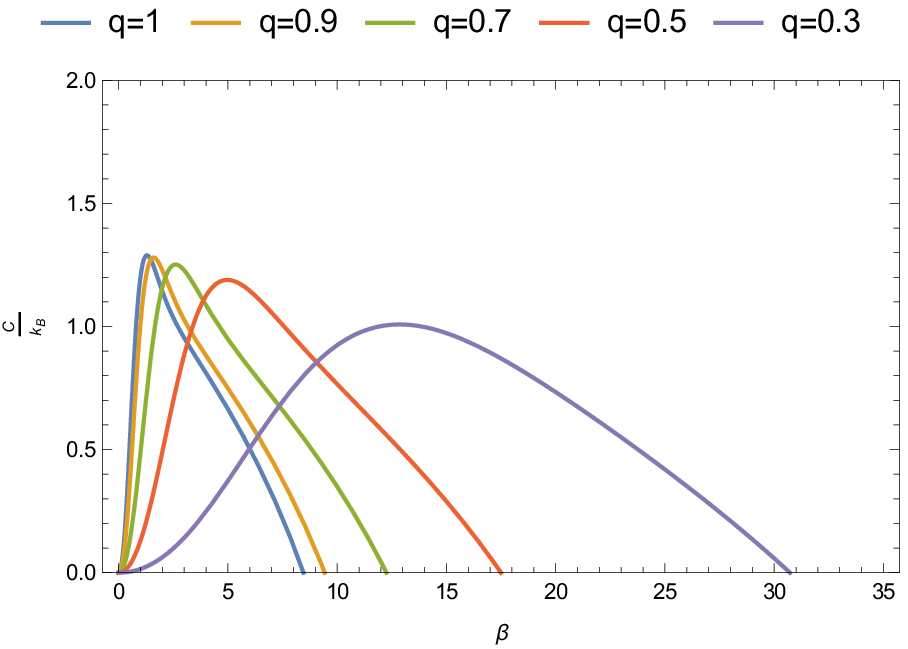}

}

\subfloat[Both vibrational partition function $Z$ and specific heat $\frac{C}{k_{B}}$
of $\text{LiH}$ as a function of $\beta$ for different values of
$q$.]{\includegraphics[scale=0.6]{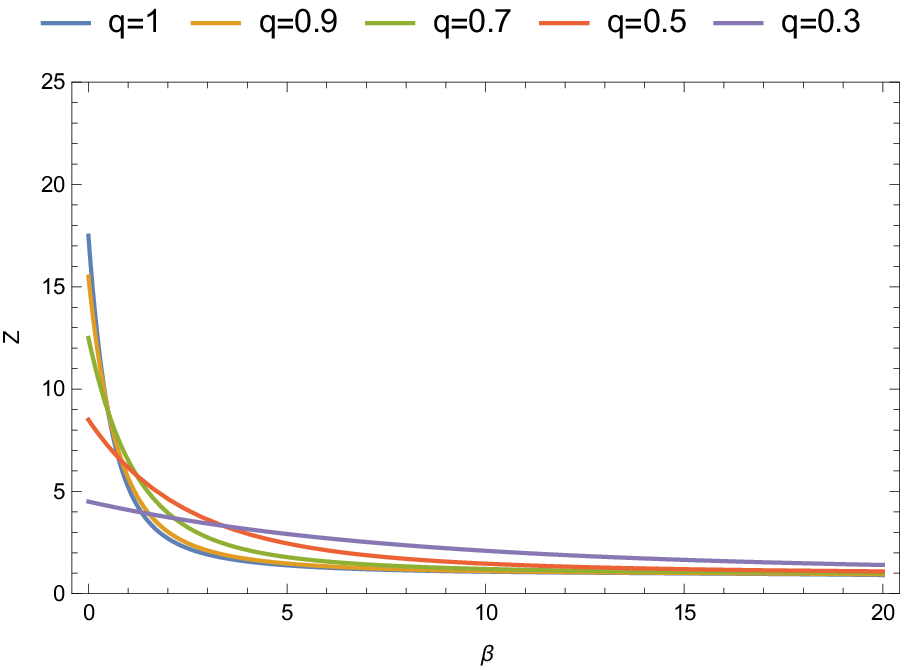}\includegraphics[scale=0.62]{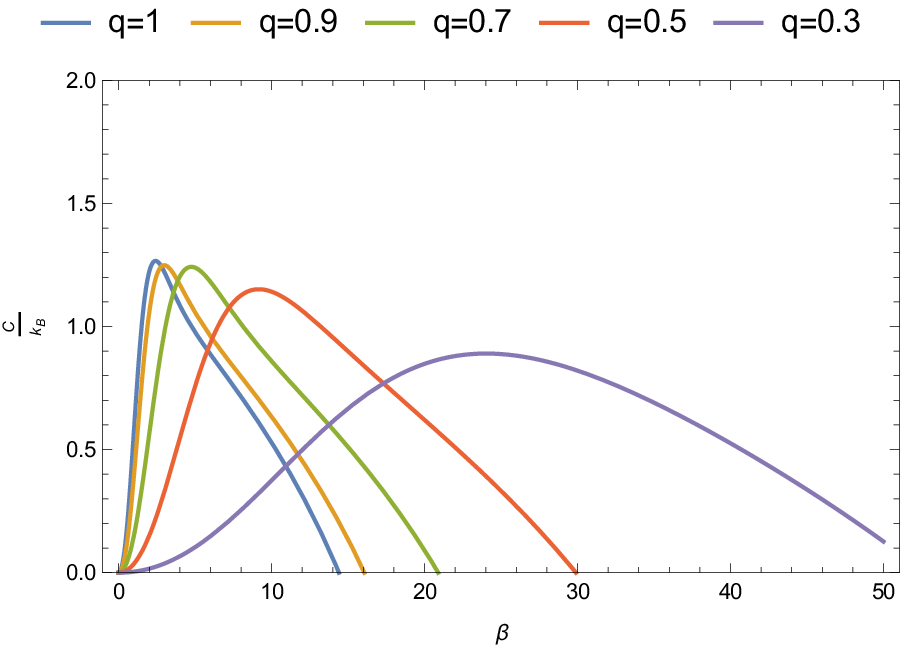}

}

\subfloat[Both vibrational partition function $Z$ and specific heat $\frac{C}{k_{B}}$
of $\text{CO}$ as a function of $\beta$ for different values of
$q$.]{\includegraphics[scale=0.6]{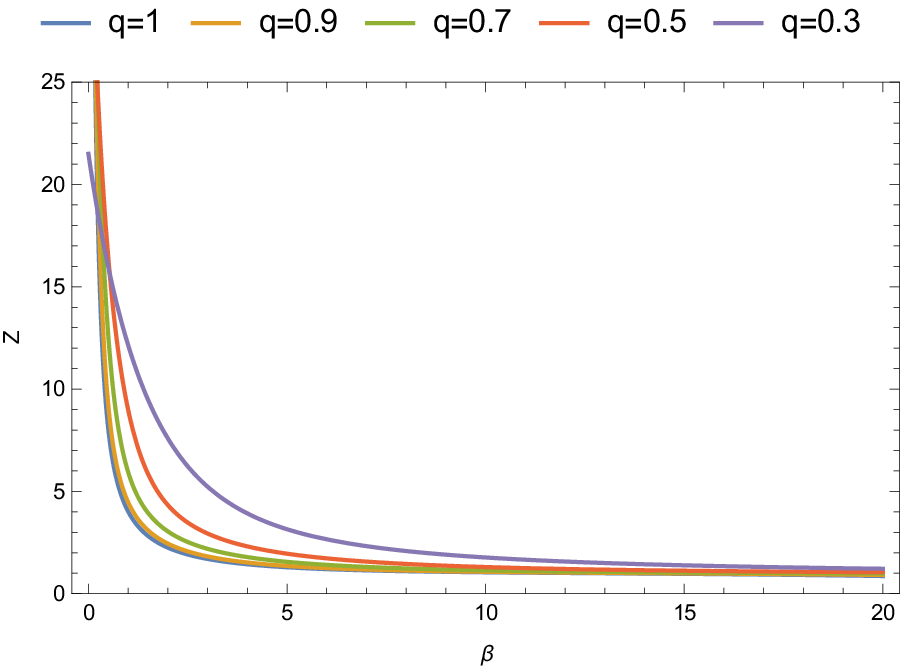}\includegraphics[scale=0.62]{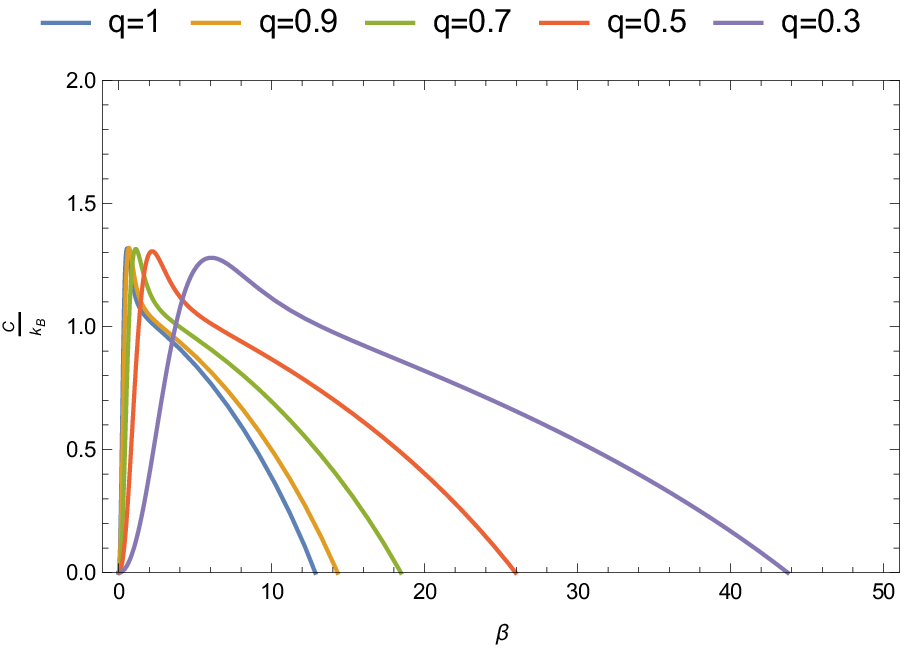}

}

\caption{\label{fig:3}q-deformed thermal properties for the following Diatomic
molecules, $\text{H}_{2}$, H Cl, LiH and CO.}

\end{figure}

The behaviour of the thermodynamic quantities is depicted in Fig. \ref{fig:3}: this figure shows both q-deformed vibrational partition function and q-deformed vibrational specific heat for the following diatomic molecules,
$\text{H}_{2}$, HCl, LiH and CO. The effects of the deformation parameter for typical values of $q$ on both partition function and specific heat are considerable. The case of standard quantum mechanics corresponding
to $q=1$ is well depicted, too in this figure. 

In Fig. \ref{fig:3}, the q-deformed partition function $Z$ vs. $\beta$
is plotted for various values of $q$ in the interval $0\le\beta\le20$.
It reveals that for increasing temperature $\beta$, the partition function is decreased as well.
The variation of heat capacity vs. $\beta$ for various values of
$q$ and in the interval $0\le\beta\le50$ is shown in Fig. \ref{fig:3}. We observe that the vibrational specific heat $\frac{C}{k_{B}}$ first increases with the increasing $\beta$ until it reaches the
maximum value $\beta_{C}$ and then decreases with it. In Table. \ref{tab:3},
we show some values of the critic temperature $T_{C}=\frac{1}{k_{B}\beta_{C}}$,
and it variation with the parameter $q$ for four diatomic molecules:
from this table, this temperature depends strongly on the parameter of deformation $q$, and increases with it. 
\begin{table}[h]
\begin{tabular}{|c|c|c|c|c|}
\hline 
$q$ & $\text{H}_{2}$ & $\text{HCl}$ & $\text{LiH}$ & $\text{CO}$\tabularnewline
\hline 
\hline 
1 & \textbf{8926} & \textbf{9512} & \textbf{5023} & \textbf{21490}\tabularnewline
\hline 
0.9 & \textbf{6826} & \textbf{7076} & \textbf{3868} & \textbf{19341}\tabularnewline
\hline 
0.7 & \textbf{4463} & \textbf{4605} & \textbf{2490} & \textbf{11604}\tabularnewline
\hline 
0.5 & \textbf{2353} & \textbf{2353} & \textbf{1289} & \textbf{5474}\tabularnewline
\hline 
0.3 & \textbf{967} & \textbf{892} & \textbf{485} & \textbf{1934}\tabularnewline
\hline 
\end{tabular}

\caption{\label{tab:3}The values of the critic temperatures $T_{C}$ for the
four diatomic molecules .}
\end{table}

In addition, we can see, the vibrational specific heat $\frac{C}{k_{B}}$
is more sensitive to $n_{max}$ than the vibrational partition function.
The reason is twofold: (i) that the specific heat depends upon the
second derivative of the partition function, and (ii) the expansion
of the specific heat as a function of $q$ is very clear in the figure
for the different types of molecules: this enlargement can be explained
by the decreases in the number of energy levels $n_{max}$ when q
decreases. This is due to the finite number of states of the algebraic
model. It means for a system composed of diatomic molecules whose
potential energy is Morse potential, it has a critical temperature
value in which the system becomes saturated and can no longer absorb
more energy because all its excited states are occupied\cite{28}.

\begin{figure}
\subfloat[$q=1$]{\includegraphics[scale=0.8]{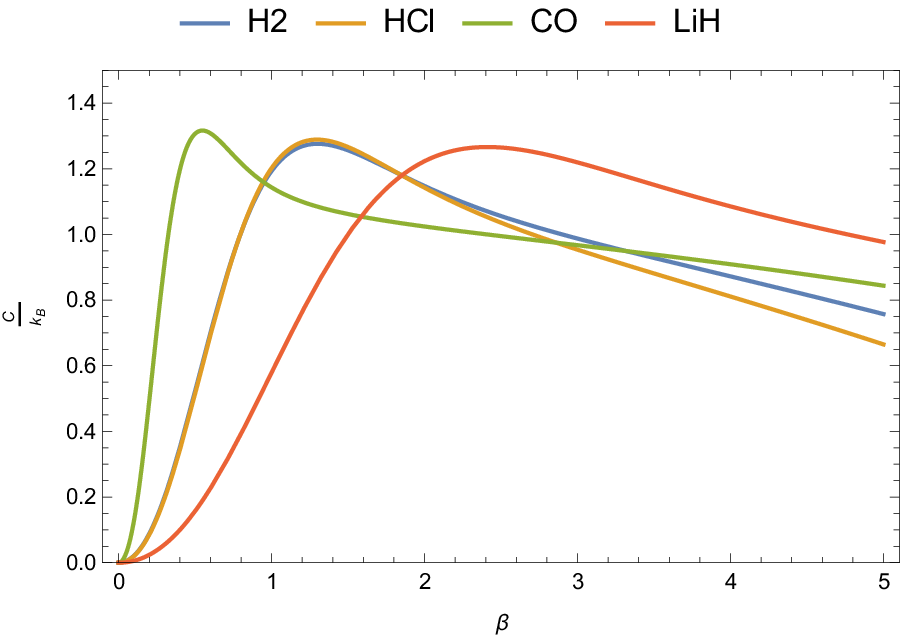}

}\subfloat[$q=0.5$]{\includegraphics[scale=0.8]{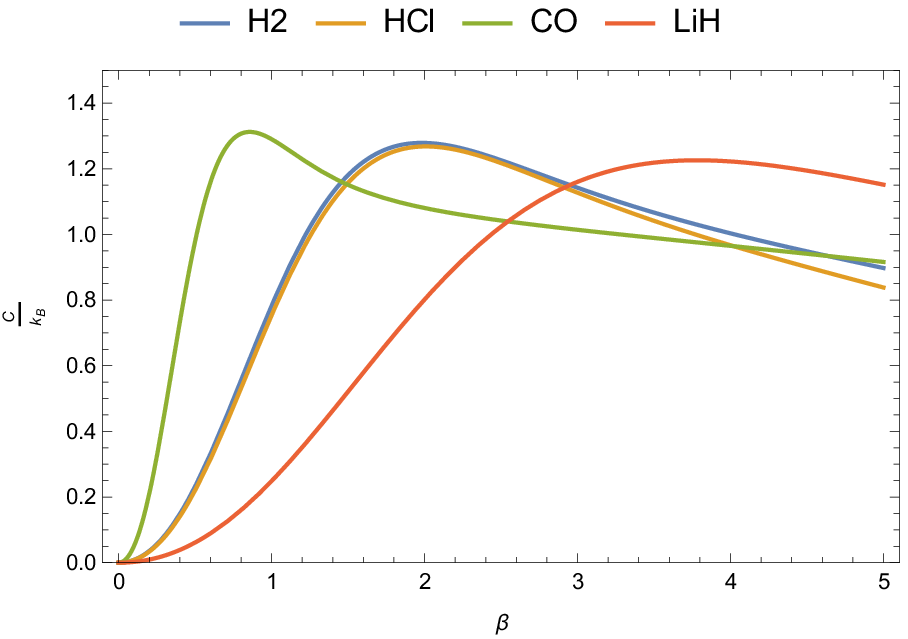}

}

\caption{\label{fig:4}q-deformed specific heat function of $\text{H}_{2}$,
HCl, LiH and CO as $\beta$ for both $q=1$ and $q=0.5$.}
\end{figure}

In Figure. \ref{fig:4}, we show the behaviour of the specific heat of diatomic molecules,$\text{H}_{2}$, HCl, LiH and CO in the same figure for two values of q: we observe that the specific heat of$\text{H}_{2}$, HCl in both cases is approximately the same, contrary compared with the case of the molecules Li H and CO where the difference is well observed. \par
\textcolor{red}{
Finally, the specific heat is an important physical quantity for testing the existence of a transition phase as well as its nature (first or second order).
Recently, predictions of entropy for diatomic molecules and gaseous substances have been the subject of two recent studies \cite{29,30}. The authors have been obtained the exact expression of the vibrational entropy for (i) the improved Tietz oscillator \cite{29}, and  (ii) for the Morse and Manning-Rosen oscillators \cite{30}. Following these authors, two remarks can be made: firstly, the diatomic molecules are treated as rigid rotors,  and  the interaction between two molecules is neglect. On the other hand, the specific heat values derived from experimental measurements are a combination of the translational, rotational and vibrational specific heat. Thus, in order to compare our theoretical vibrational specific heat, it must have the experimental data of the total specific heat at first. 
But, we stress, though, as far as we know, we have not any data concerning the experimental values of the total one-dimensional specific heat that help us to make a comparison with our theoretical results. However, this study can provide us to extend it to the three-dimensional case where this comparison perhaps can be made.
}

\section{Conclusion}

In this work, by using the vibrational energies obtained in the q-deformed Morse model, we carry out a calculation of the vibrational partition function of the Morse potential for some diatomic molecules via the Euler-Maclaurin approach. From the partition function obtained, we have derived explicit expressions for the thermodynamic functions such as specific heat C. We have analysed the behaviour of the q-deformed specific heat and introduced a critical temperature $T_{C}$ and its relation with the parameter of deformation $q$.


\begin{thebibliography}{99}
\bibitem[1]{1}D. Bonatsost, L. Britot and D. Menezes, J. Phys. A: Math. Gen. \textbf{26}, 895-904 (1993).

\bibitem[2]{2}L. C. Biedenharn, J. Phys. A: Math. Gen. \textbf{22}, L873 (1989).

\bibitem[3]{3}A. J. Macfarlane, J. Phys. A: Math. Gen. \textbf{22}, 4581 (1989).

\bibitem[4]{4}P. P. Kulish and E. V. Damaskinsky, J. Phys. A: Math. Gen. \textbf{23}, L415 (1990).

\bibitem[5]{5}Ng Y, J. Phys. A: Math. Gen. \textbf{23}, 1023 (1990)

\bibitem[6]{6}Ui H and Aizawa N 1990 Mod. Phys. Lett. A \textbf{5}, 237 (1990).

\bibitem[7]{7}A. Lorek, A. Ruffing and J. Wess, Z. Phys. C \textbf{74}, 369--377 (1997).

\bibitem[8]{8}I. L. Cooper and R. K. Gupta, Phys. Rev. A, \textbf{52}, 941 (1995).

\bibitem[9]{9}A. Lavagno, A.M. Scarfone and P. N. Swamy, Eur. Phys. J. C. \textbf{47}, 253--261 (2006).

\bibitem[10]{10}J. \v{C}rnugelj, M. Martinis and V. Mikuta-Martinis, Phys. Lett. A \textbf{188}, 347-354 (1994).

\bibitem[11]{11}M. Gavrilik, I. I. Kachurik and A.V. Lukash, Ukr.J. Phys, 58, (2013).

\bibitem[12]{12}M. S. Abdalla and H. Eleuch, J. Appl. Phys, \textbf{115}, 234906 (2014).

\bibitem[13]{13}S. Cai, G. Su and J. Chen, J. Phys. A: Math. Th,eor. \textbf{40}, 11245--11254 (2007).

\bibitem[14]{14}Won Sang Chung, Int. J. Mod. Phys B. \textbf{29}, 1550177 (2015).

\bibitem[15]{15}Won Sang Chung, Journal of Advanced Physics. \textbf{4},1--4 ( 2015).

\bibitem[16]{16}M. Angelova and A. Frank, Physics of Atomic Nuclei,\textbf{68}, 1625--1633 (2005).

\bibitem[17]{17}M. Angelova and A. Franck, Phys. Atom. Nuclei \textbf{68}, 1689-1697 (2005). 

\bibitem[18]{18}A. Algin and M. Senay, Physica A.  \textbf{447}, 232--246 (2016).

\bibitem[19]{19} M.L. Strekalov, Chem. Phys. Lett, \textbf{439}, 209-212 (2007).

\bibitem[20]{20}Chun-Sheng Jia, Lie-Hui Zhang,
Chao-Wen Wang, Chem. Phys. Lett. \textbf{667}, 211-215 (2017).

\bibitem[21]{21} \textcolor{red}{Xiao-Qin Song, Chao-Wen Wang
and Chun-Sheng Jia, Chem. Phys. Lett. \textbf{673}, 50-55 (2017).}

\bibitem[22]{22} \textcolor{red}{Chun-Sheng Jia, Chao-Wen Wang, Lie-Hui Zhang, Xiao-Long Peng, Ran Zenga and Xu-Tao You, Chem. Phys. Lett. \textbf{676}, 150-153 (2017).}

\bibitem[23]{23} G. Arfken, Mathematical Methods for Physicists, 3rd ed. (Academic Press, Orlando, FL, 1985), pp. 327--338.

\bibitem[24]{24}M. R. Setare and O. Hatam, Mod. Phys. Lett. A. \textbf{24}, 361--367 (2009). 

\bibitem[25]{25}S.-H. Dong, F. Lara-Rosana and G.-H. Sun, Phys. Lett. A. \textbf{325}, 218 (2004).

\bibitem[26]{26}I. Nasser, MS. Abdelmonem, H. Bahlouli, AD. Alhaidar. J. Phys. B. \textbf{40}:4245 (2007).

\bibitem[27]{27}A. K. Roy, Resul. in. Phys. \textbf{3}, 103--108 (2013).

\bibitem[28]{28}G. Valencia- Ortega and L. A. Arias-Hernandez, arxiv: 1708.00926v1 (2017).

\bibitem[29]{29} \textcolor{red}{J-F. Wang, X-L. Peng, L-H. Zhang, C-W. Wang and C-S. Jia	, Chem. Phys. Lett. \textbf{686}, 131-133 (2017).}

\bibitem[30]{30} \textcolor{red}{C-S. Jia, C-W. Wang, L-H. Zhanh, X-L. Peng, H-M. Tang, J-Yi. Liu, Yu. Xiong and R. Zeng, Chem. Phys. Lett. \textbf{692}, 57-60 (2018).}


\end{thebibliography}
\end{document}